\documentclass[twocolumn,aps,prc,superscriptaddress,showpacs,floatfix]{revtex4}

\usepackage{epsfig,bm,feynmf}

\def\ltap{\ \raise.3ex\hbox{$<$\kern-.75em\lower1ex\hbox{$\sim$}}\ }
\def\gtap{\ \raise.3ex\hbox{$>$\kern-.75em\lower1ex\hbox{$\sim$}}\ }

\begin{document}



\title{Charmed Exotics in Heavy Ion Collisions}


\author{Su Houng Lee}%
\email{suhoung@phya.yonsei.ac.kr}
\author{Shigehiro Yasui}
\email{yasui@phya.yonsei.ac.kr} \affiliation{Institute of Physics
and Applied Physics, Yonsei University, Seoul 120-749, Korea}
\author{Wei Liu}
\email{weiliu@comp.tamu.edu}
\author{Che Ming Ko}
\email{Ko@comp.tamu.edu} \affiliation{Cyclotron Institute and
Physics Department, Texas A\&M University, College Station, TX
77843, U.S.A.}


\begin{abstract}
Based on the color-spin interaction in diquarks, we argue that
charmed multiquark hadrons are likely to exist. Because of the
appreciable number of charm quarks produced in central
nucleus-nucleus collisions at ultrarelativistic energies, production
of charmed multiquark hadrons is expected to be enhanced in these
collisions.   Using both the quark coalescence model and the
statistical hadronization model, we estimate the yield of charmed
tetraquark meson $T_{cc}$ and pentaquark baryon $\Theta_{cs}$ in
heavy ion collisions at RHIC and LHC. We further discuss the decay
modes of these charmed exotic hadrons in order to facilitate their
detections in experiments.
\end{abstract}



\maketitle


\section{Introduction}

Possible existence of exotic mesons consisting of two quarks and two
anti-quarks was first suggested by Jaffe in the framework of the MIT
bag model \cite{Jaffe76}. Since then, there have been continuous
discussions on whether the mesons in the scalar nonet are candidates
for such tetraquark mesons. Recently, interest in tetraquark mesons
has been extended to include those containing heavy quarks
\cite{Zouzou86,Semay94}, as several heavy mesons, that were observed
in $B$ meson decays, do not seem to fit well within the conventional
quark model \cite{Zhu07}. Tetraquark mesons with two heavy
anti-quarks ($\bar{Q}\bar{Q} qq$), henceforth called $T_{QQ}$, are
particularly interesting as they are explicitly exotic from flavor
considerations \cite{Manohar93}. Moreover, simple theoretical
consideration based on the color-spin interaction \cite{DeRujula75}
shows that for such configurations the binding energy increases as
the mass of the heavy quark increases. Calculations based on the
flavor-spin interaction \cite{Glozman96,Stancu97,Stancu98} or the
instanton induced interactions \cite{Chernyshev:1995gj} also show
that the mass of $T_{cc}$ is below that of two charmed mesons. For a
similar reason, the chance of having a stable heavy pentaquark
($qqqq\bar{Q}$) increases as the mass of heavy anti-quark becomes
larger.

Experimental observation of such explicitly exotic hadrons is
crucial in refining our understanding of multiquark interactions in
low energy QCD. However, producing the $T_{QQ}$ from an elementary
process is highly suppressed as it involves creating two $\bar{Q}Q$
pairs from the vacuum.  In contrast, in relativistic heavy ion
collisions at LHC, $\bar{c}c$ pairs are expected to be abundantly
produced \cite{zhang}. Since the hadronization from the quark-gluon
plasma produced in these collisions tends to follow a statistical
description, production of exotic hadrons in heavy ion collisions at
LHC is thus much more favorable than in elementary reactions
\cite{chen1,Nonaka:2003ew, Maiani:2006ia}.

In this work, we first give a qualitative argument why multiquark
hadrons consisting of heavy quarks are likely to exist. Using both
the quark coalescence model and the statistical hadronization model,
we then give estimates of how many $T_{QQ}$ and charmed pentaquark
baryons, if they exist, will be produced in central heavy ion
collisions at both RHIC and LHC. Furthermore, possible decay modes
of these charmed exotic hadrons are discussed.

\vspace{0.5cm}
\section{A schematic model for hadron mass differences}

\subsection{known hadrons}

Sophisticated constituent quark model calculations have been
performed to study possible stable multiquark hadrons that consist
of heavy quarks. These results can be roughly understood in terms of
simple arguments based on the color-spin interaction. To illustrate
the mechanism, we introduce the following simplified form for the
color-spin interaction \cite{DeRujula75}:
\begin{eqnarray}
C_H \sum_{i>j} \vec{s}_i \cdot \vec{s_j} \frac{1}{m_i m_j}.
\label{mass1}
\end{eqnarray}
Here $m$ and $\vec{s}$ are the mass and spin of the constituent
quarks $i$ and $j$. The strength of the color-spin interaction $C_H$
should depend on the wave function and the exact form of the
interaction as well as the color structure of either the quark-quark
or quark-antiquark pair. The color factor would be $8/3$ for
diquarks in the color antitriplet channel and $16/3$ for quark and
anti-quark pair in the color singlet channel. This simple form with
$C_H=C_B$ for a diquark and $C_H=C_M$ for a quark-antiquark pair can
capture some of the essential physics in hadron masses. To
illustrate this point, we assume the following constituent quark
masses: $m_{u,d}=300~{\rm MeV}$, $m_s=500~{\rm MeV}$, $m_c=1500~{\rm
MeV}$, and $m_b=4700~{\rm MeV}$.

\begin{table}[h]
\centering
\begin{tabular}{|c|c|c|c|c|}
\hline  Mass Diff. & $M_\Delta-M_N$  & $M_\Sigma-M_\Lambda$
& $M_{\Sigma_c}-M_{\Lambda_c}$ & $M_{\Sigma_b}-M_{\Lambda_b}$
 \\[2pt] \hline Formula & $\frac{3C_B}{2 m_u^2}$
& $\frac{C_B}{m_u^2}(1-\frac{m_u}{m_s})$ &
$\frac{C_B}{m_u^2}(1-\frac{m_u}{m_c})$
&
$\frac{C_B}{m_u^2}(1-\frac{m_u}{m_b})$
 \\[2pt]
Fit   & 290 MeV   &  77 MeV  & 154 MeV  & 180 MeV    \\[2pt]
Experiment & 290 MeV & 75 MeV & 170 MeV & 192 MeV \\[2pt] \hline
\end{tabular}
\caption{Baryon mass relations.  The first column is fit to
experiments.} \label{baryon}
\end{table}

Table \ref{baryon} shows the mass differences between baryons that
are sensitive to the color-spin interaction only. By fitting $C_B$
to $M_\Delta - M_N$, we obtain ${C_B}/{m_u^2}=193$ MeV and find that
the mass differences $M_\Sigma-M_\Lambda$ and
$M_{\Sigma_c}-M_{\Lambda_c}$ are well reproduced. This is in no way
an attempt to make a best fit, but the point is that with typically
accepted constituent quark masses, the mass splitting larger than
$C_B$, reflecting that the quark and anti-quark correlation is about
3 times stronger than that between two quarks.

\begin{table}[h]
\centering
\begin{tabular}{|c|c|c|c|c|}
\hline  Mass Diff. & $M_\rho-M_\pi$ &
$M_{K^*}-M_{K}$  & $M_{D^*}-M_D$  &  $M_{B^*}-M_B$ \\[2pt] \hline
Formula & $\frac{C_M}{m_u^2}$ &  $\frac{C_M}{m_u m_s}$  &
$\frac{C_M}{m_u m_c}$  &   $\frac{C_M}{m_u m_b}$  \\[2pt]
Fit & 635 MeV   &  381 MeV  & 127 MeV    &  41 MeV \\[2pt]
Experiment & 635 MeV &  397 MeV & 137 MeV & 46 MeV  \\[2pt] \hline
\end{tabular}
\caption{Meson mass relations.  The first column is fit to
experiments. } \label{meson}
\end{table}

When both quarks are heavy, the value of  $C_H$ is expected to
become larger as the strength of the relative wave function at the
origin is substantially increased. Fitting instead its value to the
mass difference between $J/\psi$ and $\eta_c$, we find
$C_{c\bar{c}}/{m_c^2}=117$ MeV. Assuming that the corresponding
attraction between charmed diquark is three times smaller than that
between the charm quark-antiquark pair as in the case of light
quarks, we have $C_{cc}/{m_c^2}=39$ MeV.  We could introduce
additional mass dependence in $C_B$ and in $C_M$ by fitting the mass
differences in the strange, charm and bottom hadrons
 from Table I and II, respectively.
However, these introduce only minor changes in the analysis to
follow, and therefore we will just use the mass independent $C_H$'s
obtained above.

\subsection{charmed tetraquark mesons}

Using above parameters, we argue in this subsection why the doubly
charmed tetraquark meson might be stable. Let us consider a
tetraquark meson $T_{q_1q_2}$ that is made up of $ud \bar{q_1}
\bar{q_2}$. The reason we start with the $ud$ diquark is that for a
diquark the strongest attraction is expected when the two quarks are
light, and their total color, flavor and spin are all in the
antisymmetric states. Therefore, if there is any stable
configuration, it must involve a scalar $ud$ diquark.  We then add
two antiquarks in the relative $s$-wave state and look for a stable
configuration.

The stability of  $T_{q_1q_2}$ depends on whether it is
energetically favorable against recombining into two mesons of
$u\bar{q_1}$ and $d\bar{q_2}$. As we have discussed previously, the
attraction $C_M$ between a quark-antiquark pair is stronger than
$C_B$ in a diquark. This means that when both $q_1$ and $q_2$ are
light, the two-meson states would be energetically much more
favorable, and $T_{q_1 q_2}$ will not be stable. However, when $q_1$
and $q_2$ become heavy, the attraction in the quark-antiquark pair
in the meson decreases, while in $T_{q_1q_2}$ the attraction in the
$ud$ diquark remains the same and the interaction in the $\bar{q_1}
\bar{q_2}$ decreases substantially. Therefore, the tetraquark state
could become stable. A simplification in working with a spin zero
$ud$ diquark in $T_{q_1 q_2}$ is that there is no spin-spin
interaction between the $ud$ diquark and $q_1$ or $q_2$, and it is
sufficient to only estimate the attractions inside the diquark or
anti-diquark. If $q_1$ and $q_2$ are identical quarks, then their
total spin has to be zero, because their color combination is
antisymmetric in the present configuration. This means that their
total spin has to be 1, which is a repulsive combination. However,
the repulsion becomes smaller when quark masses become heavy.
Moreover, the quantum number of $T_{q_1q_2}$ has to be $1^+$, so
that it can not decay into two pseudoscalar mesons. The threshold
for its decay is then the masses of the vector and pseudo scalar
mesons.

\begin{table}[h]
\begin{center}
\begin{tabular}{c|c|c|c}
\hline \hline
$ud\bar{q}_{1}\bar{q}_{2}$ (spin=1) & $u\bar{q}_{1}$ (spin=1) & $d\bar{q}_{2}$
(spin=0) & $ud\bar{q}_{1}\bar{q}_{2}$ \\
$ - \frac{3}{4} \frac{C_{B}}{m_{u}^{2}} + \frac{1}{4}
\frac{C_{B}}{m_{q_1}^{2}} $ &
$\frac{1}{4}\frac{C_{M}}{m_{u}m_{q_1}}$ &
$-\frac{3}{4}\frac{C_{M}}{m_{u}m_{q_1}}$ &
 $ - u\bar{q}_{1} - u\bar{q}_{2}$ \\
\hline \hline
$ud\bar{s}\bar{s}$ & $K^{\ast}$ & $K$ &  \\
-127 & 92 & -285 & 63 \\
\hline
$ud\bar{c}\bar{c}$ & $D^{\ast}$ & $D$ &  \\
-143 & 31 & -95 & -79 \\
\hline
$ud\bar{b}\bar{b}$ & $B^{\ast}$ & $B$ &  \\
-145 & 10 & -30 & -124 \\
\hline
\end{tabular}
\end{center}
\caption{\small \baselineskip=0.5cm Tetraquark mesons
$ud\bar{q_1}\bar{q_2}$ with spin equal to1 for $q_1 = q_2$, where
$q_{1}$, $q_{2}=s$, $c$ and $b$. Units are in MeV.}
\label{tbl:table3}
\end{table}%

Table \ref{tbl:table3} shows the  mass difference between a
tetraquark meson with identical diquarks and the sum of vector and
pseudo scalar meson masses due to the color-spin interaction of
Eq.~(\ref{mass1}) with the $C_H$ parameters determined previously.
As expected, the mass difference decreases as $q_1$ and $q_2$ become
heavy, and the tetraquark mesons $T_{cc}$ and $T_{bb}$ with $c$ or
$b$ quarks are bound. Although our result is based on a very crude
estimate, essentially the same result has been obtained in the full
constituent quark model calculation \cite{Stancu97,Brac93} and the
QCD sum-rule calculation \cite{Navarra07}.

\begin{table}[h]
\begin{center}
\begin{tabular}{c|c|c|c}
\hline \hline
$ud\bar{q}_{1}\bar{q}_{2}$ (spin=0) & $u\bar{q}_{1}$
(spin=0) & $d\bar{q}_{2}$ (spin=0) & $ud\bar{q}_{1}\bar{q}_{2}$ \\
$ - \frac{3}{4} \frac{C_{B}}{m_{u}^{2}} -
\frac{3}{4} \frac{C_{B}}{m_{q_1} m_{q_2}}
$ & -$\frac{3}{4}\frac{C_{M}}{m_{u}m_{q_1}}$ &
$-\frac{3}{4}\frac{C_{M}}{m_{u}m_{q_2}}$ &
$ - u\bar{q}_{1} - u\bar{q}_{2}$\\
\hline \hline
$ud\bar{s}\bar{c}$ & $K$ & $D$ &  \\
-162 & -285 & -95 & 218 \\
\hline
$ud\bar{s}\bar{c}$ & $K$ & $B$ &  \\
-150 & -285 & -30 & 165 \\
\hline
$ud\bar{c}\bar{b}$ & $D$ & $B$ &  \\
-146 & -95 & -30 & -21 \\
\hline
\end{tabular}
\end{center}
\caption{\small \baselineskip=0.5cm Tetraquark mesons
$ud\bar{q_1}\bar{q_2}$ with spin equal to 0 for $q_1 \ne q_2$.
$q_{1}$, $q_{2}=s$, $c$ and $b$. Units are in MeV.}
\label{tbl:table4}
\end{table}%

\begin{table}[h]
\begin{center}
\begin{tabular}{c|c|c|c}
\hline \hline
$ud\bar{q}_{1}\bar{q}_{2}$ (spin=1) & $u\bar{q}_{1}$
(spin=1) & $d\bar{q}_{2}$ (spin=0) & $ud\bar{q}_{1}\bar{q}_{2} $ \\
$ - \frac{3}{4} \frac{C_{B}}{m_{u}^{2}} + \frac{1}{4}
\frac{C_{B}}{m_{q_1} m_{q_2}} $ & $\frac{1}{4}\frac{C_{M}}{m_{u}m_{q_1}}$
& $-\frac{3}{4}\frac{C_{M}}{m_{u}m_{q_2}}$ & $ - u\bar{q}_{1} -
u\bar{q}_{2}$\\
\hline
\hline
 & $K^{\ast}$ & $D$ &  \\
$ud\bar{s}\bar{c}$  & 95 & -95 & -139 \\
\cline{2-4}
-139 & $D^{\ast}$ & $K$  & \\
 &  31 & -285 & 114 \\
\hline
 & $K^{\ast}$ & $B$ &  \\
$ud\bar{s}\bar{b}$  & 95 & -30 & -208 \\
\cline{2-4}
-143 &  $B^{\ast}$ & $K$ & \\
 & 10 & -285 &  132 \\
\hline
 & $D^{\ast}$ & $B$ &  \\
$ud\bar{c}\bar{b}$  & 31 & -30 & -145 \\
\cline{2-4}
-144 & $B^{\ast}$ & $D$ &  \\
 & 10 & -95 & -59 \\
\hline
\end{tabular}
\end{center}
\caption{\small \baselineskip=0.5cm Tetraquark mesons
$ud\bar{q_1}\bar{q_2}$ with spin equal to 1 for $\bar{q}_{1} \neq
\bar{q}_{2}$, where $q_{1}$, $q_{2}=s$, $c$ and $b$. Units are in
MeV.} \label{tbl:table5}
\end{table}%

For $q_1$ and $q_2$ of different flavors, their total spin could be
either zero or one. The quantum number of the tetraquark meson could
then be either $0^+$ or $1^+$. Table \ref{tbl:table4} and Table
\ref{tbl:table5} show the mass differences in such cases. As in the
previous case, bound tetraquark mesons with $\bar{c}\bar{b}$ could
exist.

\subsection{charmed pentaquark baryons}

\begin{table}[h]
\begin{center}
\begin{tabular}{c|c|c|c}
\hline \hline
$udud\bar{q}$ & $uud$ & $d\bar{q}_{2}$ & $udud\bar{q} - uud - d\bar{q}$ \\
 $ 2 \left( - \frac{3}{4} \frac{C_{B}}{m_{u}^{2}} \right) +
 \Delta E_{L=1}$ & $-\frac{3}{4}\frac{C_{M}}{m_{u}^{2}}$ &
 $-\frac{3}{4}\frac{C_{M}}{m_{u}m_{q}}$ & \\
\hline \hline
$udud\bar{s}$ & $N$ & $K$ &  \\
-290+$\Delta E_{L=1}$ & -145 & -286 & 141+$\Delta E_{L=1}$ \\
\hline
$udud\bar{c}$ & $N$ & $D$ &  \\
-290+$\Delta E_{L=1}$ & -145 & -95 & -50+$\Delta E_{L=1}$ \\
\hline
$udud\bar{b}$ & $N$ & $B$ &  \\
-290+$\Delta E_{L=1}$ & -145 & -30 & -114+$\Delta E_{L=1}$ \\
\hline
\end{tabular}
\end{center}
\caption{\small \baselineskip=0.5cm Charm and bottom pentaquark
baryons $\Theta_{q}(udud\bar{q})$ ($q=c$ and $b$) with spin equal to
1. $\Delta E_{L=1}=309$ MeV is en excitation energy of two diquarks
with relative angular momentum $L=1$. Units are in MeV.}
\label{tbl:table6}
\end{table}%

Similar observations can be made for heavy pentaquark baryons. Many
constituent quark model calculations show that the $\Theta^+$
\cite{Leps03}, if it exists at all, can not be explained as a bound
state of $udud\bar{s}$ constituent quarks \cite{Hiyama:2005cf}. This
is due to the strong attraction between the $\bar{s}$ and the light
quark, so that it is energetically much more favorable for
$udud\bar{s}$ to form a meson and a baryon. The attraction to form a
meson becomes smaller if the $\bar{s}$ is replaced by either a
$\bar{c}$ or $\bar{b}$.   Full constituent quark model calculations
\cite{Stancu:2004du,Stancu:2005jv} indeed find a possible stable
heavy pentaquark baryon. A likely pentaquark structure would be that
suggested in Ref.~\cite{JW03} with the two scalar diquark $ud$
combined into an $L=1$ and color antisymmetric state. The excitation
energy of a diquark in a $L=1$ state, $\Delta E_{L=1}$, can be
estimated by approximating the charmed baryon as a sum of a charm
quark and a diquark, because the interaction between them is small
in the heavy quark limit. Attributing the mass difference between
the parity doublet partners of the positive parity $\Lambda_{c}^{+}$
(2286 MeV) and the negative parity $\Lambda_{c}^{\ast +}$ (2595 MeV)
to the $L=1$ excitation of the diquark, as the heavy charm quark
would act as the center of mass, leads to $\Delta E_{L=1}=309$ MeV.
Applying this $L=1$ excitation energy to the relative excitation of
two diquarks, we find that while a strange pentaquark baryon is very
unlikely to exist, heavy pentaquark baryons $\Theta_c$ and
$\Theta_b$ could be closer to the threshold  as shown in Table
\ref{tbl:table6}, consistent with the full constituent quark model
calculation \cite{Stancu:2004du,Stancu:2005jv,Maltman04} and the QCD
sum-rules study \cite{SKL05} in which a possible stable heavy
pentaquark baryon has been found.

\begin{table}[h]
\begin{center}
\begin{tabular}{c|c|c|c}
\hline \hline
 & $N$ & $s\bar{q}$ & $udus\bar{q} - N - s\bar{q}$ \\
 & $-\frac{3}{4}\frac{C_{M}}{m_{u}^{2}}$ & $-\frac{3}{4}
 \frac{C_{M}}{m_{u}m_{q}}$ &  \\
\cline{2-4} $udus\bar{q}$ & $\Sigma$ & $d\bar{q}$ & $udus\bar{q} -
\Sigma -
d\bar{q}$ \\
$ - \frac{3}{4} \frac{C_{B}}{m_{u}^{2}} - \frac{3}{4}
\frac{C_{B}}{m_{u} m_{s}} $ & $ \frac{1}{4} \frac{C_{B}}{m_{u}^{2}}
-\frac{C_{B}}{m_{u}m_{s}}$ & $-\frac{3}{4}\frac{C_{M}}{m_{u}m_{q}}$ & \\
\cline{2-4}
 & $\Lambda$ & $u\bar{q}$ & $udus\bar{q} - \Lambda - u\bar{q}$ \\
& $ -\frac{3}{4} \frac{C_{B}}{m_{u}^{2}} $ & $-\frac{3}{4}
\frac{C_{M}}{m_{u}m_{q}}$ & \\
\hline \hline
 & $N$ & $D_{s}$ & $udus\bar{c} - N - D_{s}$ \\
 & -145 &  -57 & -30 \\
 \cline{2-4}
 $udus\bar{c}$ & $\Sigma$ & $D$ & $udus\bar{c} - \Sigma - D$ \\
 -232 & -67 & -95 & -69 \\
 \cline{2-4}
  & $\Lambda$ & $D$ & $udus\bar{c} - \Lambda - D$ \\
 & -145 & -95 & 8 \\
\hline
 & $N$ & $B_{s}$ & $udus\bar{b} - N - B_{s}$ \\
 & -145 &  -18 & -68 \\
 \cline{2-4}
  $udus\bar{b}$ & $\Sigma$ & $B$ & $udus\bar{b} - \Sigma - B$ \\
 -232 & -67 & -30 & -133 \\
  \cline{2-4}
    & $\Lambda$ & $B$ & $udus\bar{c} - \Lambda - B$ \\
 & -145 & -30 & -56 \\
\hline
\end{tabular}
\end{center}
\caption{\small \baselineskip=0.5cm Charm- and bottom-strange
pentaquark baryons $\Theta_{qs}(udus\bar{q})$ ($q=c$ and $b$) with
spin equal to 0. Units are in MeV.} \label{tbl:table7}
\end{table}%

For a pair of $ud$ and $us$ diquarks in $\Theta_{cs}(udus\bar{c})$,
they do not have to be in the $L=1$ state, and hence there is no
additional contribution from the orbital energy \cite{Lipkin87}. The
result from our simple estimates are given in table
\ref{tbl:table7}. Previous experiments
\cite{Aitala:1997ja,Aitala:1999ij} have tried to search for this
pentaquark baryon assuming that it is bound and has a lifetime
similar to that of $D_s$. The experiment could only determine an
upper bound of greater than 0.02 for its production cross section
relative to that for the $D_s$, which is larger than typical
theoretical estimates. From simple application of statistical
hadronization model, the number of $\Theta_{\bar{c}s}$ to that of
$D_s$ is roughly $\exp(-(m_{\Theta_{\bar{c}s}}-m_{D_s})/T) \sim
\exp(-5)=0.007$, assuming a hadronization temperature of $T=200$
MeV. This is smaller than the experimental upper bound and therefore
further search is essential.

\vspace{0.5cm}
\section{production of charmed exotics in relativistic heavy ion collisions}

As discussed above, tetraquark mesons and pentaquark baryons are
more likely to exist in the heavy quark sector such as the $T_{cc}$,
$T_{cb}$, $T_{bb}$, $\Theta_{cs}$, and $\Theta_c$. It is, however,
very unlikely that they can be observed in $B$ decays or elementary
processes, as the favorable exotics involve two heavy quarks.
However, the abundance of heavy quarks is significantly enhanced in
ultrarelativistic heavy ion collisions, e.g. in a central collision
at the LHC, more than 20 $c\bar{c}$ pairs are expected to be
produced in one unit of midrapidity. Therefore, while a heavy quark
produced in an elementary process will most likely find a heavy
antiquark instead of a heavy quark, the probability to find a heavy
antiquark or a heavy quark in a heavy ion collision is similar. The
probability to form a $T_{cc}$ compared to a $J/\psi$ thus will only
be suppressed by the additional statistical factor coming from
combining an additional $ud$ diquark.

\vspace{0.5cm}
\subsection{charmed tetraquark mesons}

The number of heavy tetraquark mesons produced from the quark-gluon
plasma formed in relativistic heavy ion collisions can be estimated
in the coalescence model \cite{Mattie95}, which has been shown
to describe very well the pion and proton transverse momentum
spectra at intermediate momenta \cite{hwa,fries} as well as at low
momenta if resonances are included \cite{greco}, and the yield and
transverse momentum spectra of phi meson and Omega baryon
\cite{chen2} as well as the charmed meson \cite{greco1}. We employ
the formula that was previously used to calculate the yields of
tetraquark $D_{sJ}$(2317) meson \cite{chen} and pentaquark
$\Theta^+$ baryon \cite{chen1} at RHIC to study $T_{cc}$ production
in central Au+Au collisions at RHIC and Pb+Pb collisions at LHC. In
this model, the $T_{cc}$ number is given by
\begin{eqnarray}\label{coalmodel}
N_{T_{cc}}^{\rm coal} &=&g_{T_{cc}}\int_{\sigma_C}
\prod_{i=1}^{4}\frac{p_{i}\cdot d\sigma
_{i}d^{3}\mathbf{p}_{i}}{(2\pi )^{3}E_{i}}f_{q}(x_{i},p_{i})
\nonumber \\
&&\times f_{T_{cc}}^W(x_{1}..x_{4};p_{1}..p_{4}).
\label{coal}
\end{eqnarray}
In the above, the color-spin-isospin factor
$g_{T_{cc}}=3\times1/3^4\times1/2^4=1/432$ is the color-spin-isospin
factor for the four quarks to form a hadron of the quantum number of
the tetraquark meson and $d\sigma$ denotes an element of a
space-like hypersurface  at hadronization. Assuming Bjorken
correlation $y=\eta$ between the space-time rapidity $\eta$ and the
momentum-energy rapidity $y$ and neglecting the transverse flow as
well as using the non-relativistic approximation, we obtain the
following expression for the number of $T_{cc}$ produced from quark
coalescence:
\begin{eqnarray}\label{fourquark}
N_{T_{cc} }&\simeq& \frac{1}{432} \frac{N_{\bar c}N_{\bar
c}N_uN_d}{2}\prod_{i=1}^3\frac{(4\pi\sigma_i^2)^{3/2}}{V_C(1+2\mu
_{i}T_C\sigma_{i}^{2})},
\end{eqnarray}
where $T_C=170$ MeV is the critical temperature and $V_C$ is the
fireball volume at hadronization, which is about 1,000 fm$^3$ in
central Au+Au collisions at $s^{1/2}_{NN}=200~{\rm GeV}$ \cite{chen}
and about 2,700 fm$^3$ in central Pb+Pb collisions at
$s^{1/2}_{NN}=5.5~{\rm TeV}$ \cite{zhang}. The quark numbers at
hadronization are denoted by $N_u$ and $N_d$ for  light quarks and
$N_c$ and $N_{\bar c}$ for  heavy quarks. Their values are taken to
be $N_u=N_d=245$ \cite{chen} and 662 \cite{zhang} as well as
$N_c=N_{\bar c}=3$ and 20 in central RHIC and LHC collisions,
respectively, all in one unit of midrapidity. The charm quark
numbers are based on initial hard scattering of nucleons in the
colliding nuclei \cite{chen,zhang}. In obtaining
Eq.~(\ref{fourquark}), we have used the quark momentum distribution
function
\begin{eqnarray}
f_q(x,p)=6\delta(\eta-y){\rm exp}(-(m_q^2+p_T^2)^{1/2}/T_c)
\end{eqnarray}
and the tetraquark meson Wigner distribution function
\begin{equation}\label{wigner2}
f_{T_{cc}}^{W}(x;p)=8^3\exp \left(-\sum_{i=1}^{3}
\frac{\mathbf{y}_{i}^{2}}{\sigma_{i}^{2}}
-\sum_{i=1}^{3}\mathbf{k}_{i}^{2}\sigma_{i}^{2}\right),
\end{equation}
where the relative coordinates $\mathbf{y}_{i}$ and momenta
$\mathbf{k}_{i}$ are related to the quark coordinates
$\mathbf{x}_{i}$ and momenta $\mathbf{p}_{i}$ by the Jacobian
transformations defined in Eqs.~(7) and (8) of Ref.~ \cite{chen}.
The width parameter $\sigma_i$ in the Wigner function is related to
the oscillator frequency $\omega$ by $\sigma_i=1/\sqrt{\mu_i\omega}$
with the reduced masses $\mu_i$ defined in Eq.~(9) of
Ref.~\cite{chen}.

\begin{figure}[h]
\begin{center}
\includegraphics[height=2.2in, keepaspectratio, angle=0]{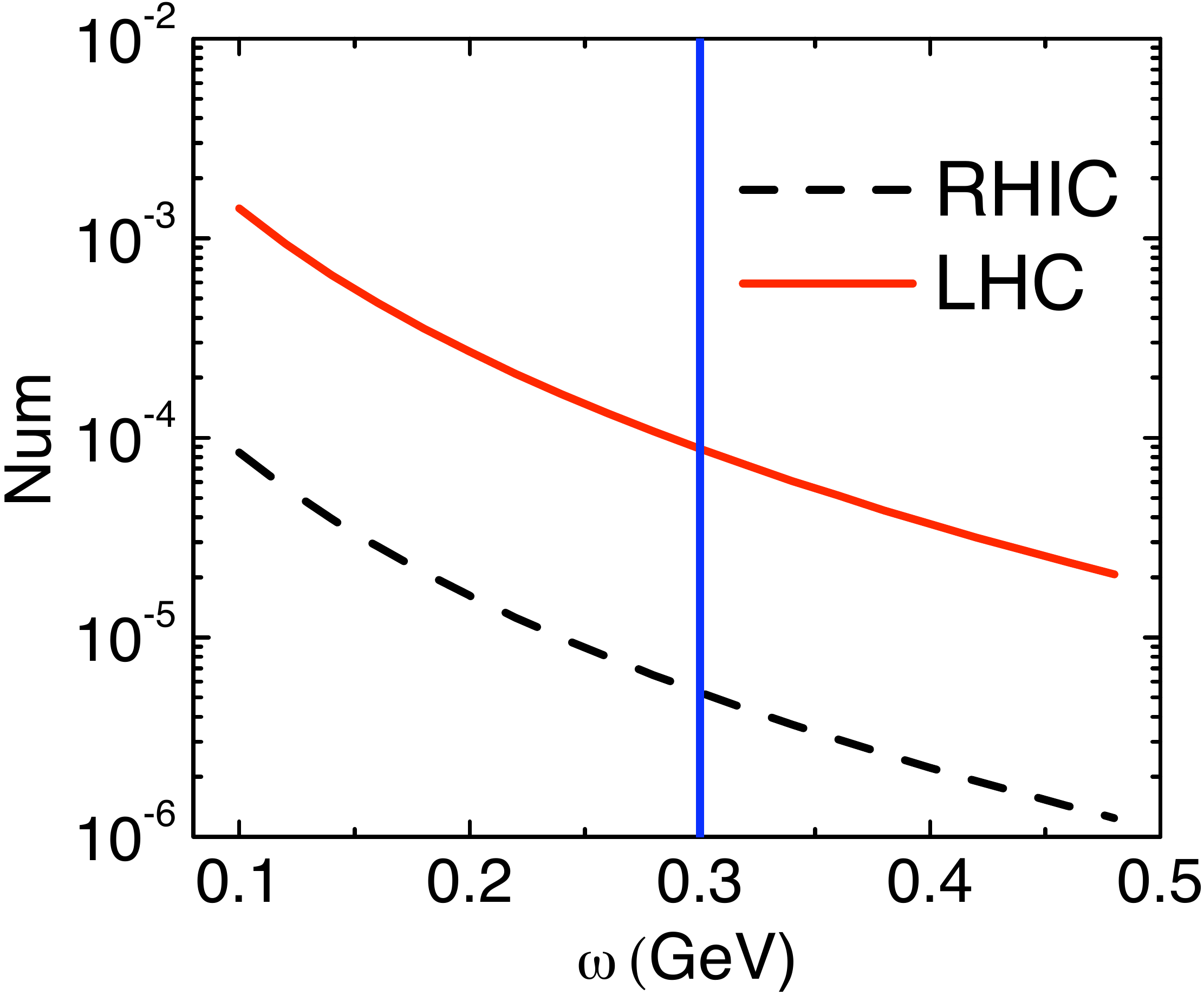}
\end{center}
\caption{Numbers of $T_{cc}$ produced at RHIC and LHC as
functions of the oscillator frequency used for the quark wave
functions in $T_{cc}$}. \label{number}
\end{figure}

In Fig.~\ref{number}, we show the numbers of $T_{cc}$ produced at
RHIC and LHC as functions of the oscillator frequency. Because of
the larger abundance of charm quarks at LHC than at RHIC, the number
of $T_{cc}$ produced at LHC is more than an order of magnitude
larger than that produced at RHIC. For the oscillator frequency
$\omega=0.3$ GeV, determined from the size $\langle
r_{D_s}^2\rangle_{\rm ch}\approx 0.124~{\rm fm}^2$ of the
$D_s^+(c\bar s)$ meson based on the light-front quark model
\cite{hwang}, the number of $T_{cc}$ produced at RHIC and LHC is
about  $5.5 \times 10^{-6}$ and $9.0 \times 10^{-5}$, respectively.

It is of interest to compare the predicted number of $T_{cc}$ mesons
from the coalescence model with that from the statistical model.  In
this model the number of $T_{cc}$ mesons produced at hadronization
is given by \cite{chen}:
\begin{eqnarray}\label{statistical}
N_{T_{cc}}^{\rm stat} &\approx&\frac{V_H\gamma_C^2 }{(2\pi)^2}\int
dm_T m_{T}^{2}e^{-\frac{{\bar\gamma}_H m_T}{T_H}}I_{0}
\left(\frac{{\bar\gamma}_H{\bar\beta}_H p_{T}}{T_C}\right),
\end{eqnarray}
where $V_H$ and ${\bar\beta}_H$ are the volume and radial flow
velocity of formed hadronic matter, and $\gamma_C$ is the fugacity
parameter for ensuring that the number of charmed hadrons produced
statistically at hadronization is same as the number of charm quarks
in the quark-gluon plasma. With $V_H\approx 1,908~{\rm fm}^3$,
$T_H=175~{\rm MeV}$, ${\bar\beta}_H=0.3c$, and the charm fugacity
$\gamma_C\approx 8.4$ \cite{chen}, we obtain $N_{T_{cc}}\sim
7.5\times 10^{-4}$ in central Au+Au collisions at RHIC. The yield of
$T_{cc}$ increases to $8.6\times 10^{-3}$ in central Pb+Pb
collisions at LHC where we have used $V_H\approx 5,220~{\rm fm}^3$,
$T_H=175~{\rm MeV}$, ${\bar\beta}_H=0.47c$, and the charm fugacity
$\gamma_C\approx 16.3$ \cite{zhang}. Compared to those from the
coalescence model, predictions from the statistical model are almost
two orders of magnitude larger.

\vspace{0.5cm}
\subsection{charmed pentaquark baryons}

For the yield of pentaquark baryon $\Theta_{cs}(udus\bar c)$,
the coalescence model gives
\begin{eqnarray}
\label{tetraquark}
N_{\Theta_{cs} }&\simeq& \frac{1}{3888}N_{\bar c}
\frac{N_s N_u N_u N_d}{2}\prod_{i=1}^4
\frac{(4\pi\sigma_i^2)^{3/2}}{V_C(1+2\mu
_{i}T_C\sigma_{i}^{2})}.
\end{eqnarray}
Using again the oscillator frequency $\omega=0.3$ GeV and taking the
anti-strange quark numbers to be 150 \cite{chen} and 405
\cite{zhang} at  RHIC and LHC, respectively, the numbers of
$\Theta_{cs}$ produced at RHIC and LHC are about $1.2\times10^{-4}$
and $7.9\times 10^{-4}$, respectively.

Since the predicted numbers of $D_s$ meson from the coalescence
model are about $5.3\times 10^{-2}$ at RHIC and $0.58$ at LHC, the
estimated ratio of numbers of $\Theta_{cs}$ and $D_{s}$ is about
$2.3 \times 10^{-3}$ at RHIC and LHC. This is consistent with the
Boltzmann factor due to the $uud$ component in $\Theta_{cs}$. In
fact, extracting $s\bar{c}$ component from $uuds\bar{c}$, the
remaining $uud \sim N$ component has a Boltzmann factor $e^{-m_{N} /
T} \simeq 4.0 \times 10^{-3}$ with $T=170$ MeV. Similar estimate
also works for the case of the $\Lambda$ and $D$ in $\Theta_{cs}$.
Using the value $0.16$ and $1.1$ for the $D$ meson numbers at RHIC
and LHC, respectively, the calculated ratio of numbers between
$\Theta_{cs}$ and $D$ is about $0.74 \times 10^{-3}$, while the
$uds$ component has a Boltzmann factor of $e^{-m_{\Lambda} / T}
\simeq 1.4 \times 10^{-3}$.

In the statistical model, the yield of $\Theta_{\bar{c}s}$ is given
by a formula similar to Eq.~(\ref{statistical}) except the power in
the charm fugacity parameter $\gamma_C$. Since there is only one
charm quark in $\Theta_{cs}$, the yield is only proportional to
$\gamma_C$. Using same parameters for evaluating the yield of
$T_{cc}$, we obtain $4.5\times 10^{-3}$ and $2.7\times 10^{-2}$ for
$\Theta_{cs}$ produced in central Au+Au collisions at RHIC and
central Pb+Pb collisions at LHC, respectively. These values are
again significantly larger than those predicted from the coalescence
model.

\vspace{0.5cm}
\section{Decay modes of charmed exotics}

In this section, we discuss the observable decay modes of the
tetraquark $T_{cc}$ and the pentaquark $\Theta_{cs}$. As we have
discussed already, $T_{cc}$ is most likely a stable state, since its
mass is below the threshold of $D^{\ast}D$. To be more general, we
consider nevertheless both cases where the mass of $T_{cc}$ is above
or below the threshold, and discuss in each case possible decay
modes that can be realistically detected in experiments with good
performance. For the $T_{cc}$ above the threshold of $D^{\ast}D$, it
can decay to $D^{\ast -}\bar{D}^{0}$ via a strong process
\footnote{The decay to the $\bar{D}^{\ast 0} D^{-}$ mode may not be
a good signal in experiments, since the $\bar{D}^{\ast 0}$ decays to
$\bar{D}^{0}\pi^{0}$ instead to $D^{+}\pi^{-}$ and $D^{-}\pi^{+}$,
which are energetically  forbidden due to mass difference.}. For the
$T_{cc}$ below the threshold of $D^{\ast}D$ and above $DD\pi$, the
decay channel to $D^{\ast -}\bar{D}^{0}$ is energetically forbidden,
but the $D^{\ast -}$ component in $T_{cc}$ can decay through a
strong process, leading to the final decay mode
$\bar{D}^{0}\bar{D}^{0}\pi^{-}$. On the other hand, when $T_{cc}$ is
below the threshold of $DD\pi$, the decay channel of $D^{\ast -}$ is
closed and only the weak decay of $\bar{D}^{0}$ component in
$T_{cc}$ is allowed via $\bar{D}^{0} \rightarrow K^{+}\pi^{-}$ or
$K^{+}\pi^{+}\pi^{-}\pi^{-}$. Therefore, $T_{cc}$ would be detected
by the decay modes $D^{\ast -} K^{+}\pi^{-}$ and $D^{\ast -}
K^{+}\pi^{+}\pi^{-}\pi^{-}$. The last two decay patterns would most
likely happen  since the binding energy of $T_{cc}$ is estimated to
be about 80 MeV as shown previously, which is sufficiently larger
than the mass difference (about 6 MeV) between $D^{\ast -}$ and
$\bar{D}^{0}\pi^{-}$. Below the threshold of $DD\pi$, it may be also
interesting to see the decay of $D^{\ast -}$ component in $T_{cc}$.
Considering that $D^{\ast -}$ component contains a quantum number of
$\bar{D}^{0}\pi^{-}$, and $\bar{D}^{0}$ decays into $K^{+}\pi^{-}$
and $K^{+}\pi^{+}\pi^{-}\pi^{-}$, we may observe the
$\bar{D}^{0}K^{+}\pi^{+}\pi^{-}$ and
$\bar{D}^{0}K^{+}\pi^{+}\pi^{+}\pi^{-}\pi^{-}$ decays.

Among the weak decays below the threshold of $DD\pi$, the decay of
the $\bar{D}^{0}$ component in $T_{cc}$ can be distinguished from
that of the $D^{\ast -}$ component. The former has the correlations
$(K^{+}\pi^{-})(K^{+}\pi^{-})\pi^{-}$ and
$(K^{+}\pi^{+}\pi^{+}\pi^{-})(K^{+}\pi^{-})\pi^{-}$, and the latter
has the correlations $(K^{+}\pi^{-})(K^{+}\pi^{+}\pi^{-})$ and
$(K^{+}\pi^{-})(K^{+}\pi^{+}\pi^{-}\pi^{-}\pi^{-})$, where brackets
denote correlated particles. However, the
$\bar{D}^{0}\bar{D}^{0}\pi^{-}$ state, which would appear in
$T_{cc}$ in the latter process, contains six quarks, hence further
analysis is needed to discuss its stability.

The pentaquark $\Theta_{cs}$ also has interesting decay patterns. As
can be seen in table \ref{tbl:table7}, the mass of $\Theta_{cs}$
could be slightly above the $\Lambda \bar{D}^0$ threshold, in which
case its lifetime will be shorter than that of $D_s$.  Then, the
only possible way to look for it is from the hadronic decay to
$\Lambda + \bar{D}^0$ final states.  Since ALICE will be able to
reconstruct the $\bar{D}^0$ through its hadronic decay, it will be
an excellent opportunity to search for $\Theta_{cs}$. Considering
more general cases, and assuming $\Theta_{cs}$ to be above the
threshold of $ND_{s}$, the $\Theta_{cs}$ can decay into $p
D_{s}^{-}$ and $\Lambda \bar{D}^{0}$ or $\Lambda D^{-}$ via the
strong process. Although the $\Sigma D$ channel is also a possible
decay mode, it is more difficult to detect as compared to $N D_{s}$
and $\Lambda D$. When the mass of $\Theta_{cs}$ is below the
$ND_{s}$ and above the $\Lambda D$ threshold, it decays only to
$\Lambda \bar{D}^{0}$ or $\Lambda D^{-}$. On the other hand, below
the threshold of $\Lambda D$, the hadronic decay channels are closed
and only weak decays are possible. In this case, the lifetime of
$\Theta_{cs}$ will depend on the lifetime of the different
components inside the $\Theta_{cs}$, such as the $\Lambda$,
$\bar{D}^{0}$ and $D^{-}$, whose lifetimes are respectively $2.6
\times 10^{-10}$, $0.41 \times 10^{-12}$ and $1.0 \times 10^{-12}$
seconds. Therefore, once the $\Theta_{cs}$ is formed as a deeply
bound state, it will decay by the weak process of $\bar{D}^{0}$ or
$D^{-}$. Consequently, possible final states would be $\Lambda
K^{+}\pi^{-}$, $\Lambda K^{+}\pi^{+}\pi^{-}\pi^{-}$ and $\Lambda
K^{+}\pi^{-}\pi^{-}$.

Since the lifetimes of $T_{cc}$ and $\Theta_{cs}$  are in the order
of $10^{-12}$ seconds, their decays occur outside the collision
region and they are thus identifiable by vertex reconstruction.
Therefore, $T_{cc}$ and $\Theta_{cs}$ would be identified clearly in
experiments if they exist. We summarize our results on possible
decay modes of $T_{cc}$ and $\Theta_{cs}$ in Tables \ref{table10}
and \ref{table11}.

\begin{widetext}
\begin{table}[h]
\caption{Possible decay modes of $T_{cc}$. In the bottom column, we
would observe correlations $(K^{+}\pi^{-})(K^{+}\pi^{-})\pi^{-}$ and
$(K^{+}\pi^{+}\pi^{+}\pi^{-})(K^{+}\pi^{-})\pi^{-}$ in the final
states. See the text for details.}
\begin{center}
\begin{tabular}{c|c|c}
\hline
  threshold & decay mode & lifetime \\
\hline
$M_{T_{cc}} > M_{D^{\ast}}+M_{D}$ & $D^{\ast -}\bar{D}^{0}$ & hadronic decay \\
$2M_{D}+M_{\pi} < M_{T_{cc}} < M_{D^{\ast}}+M_{D}$ & $\bar{D}^{0}
\bar{D}^{0}\pi^{-}$ & hadronic decay \\
$M_{T_{cc}} < 2M_{D}+M_{\pi}$ & $D^{\ast -} K^{+} \pi^{-}$,
$D^{\ast -} K^{+} \pi^{+} \pi^{-} \pi^{-}$ & $0.41 \times 10^{-12}$ sec. \\
\hline
\end{tabular}
\end{center}
\label{table10}
\end{table}

\begin{table}[h]
\caption{Possible decay modes of $\Theta_{cs}$.}
\begin{center}
\begin{tabular}{c|c|c}
\hline
  threshold & decay mode & lifetime \\
\hline
$M_{\Theta_{cs}} > M_{N}+M_{D_{s}}$ & $pD_{s}^{-}$ & hadronic decay \\
$M_{\Lambda}+M_{D} < M_{\Theta_{cs}} < M_{N}+M_{D_{s}}$ & $\Lambda
\bar{D}^{0}$ & hadronic decay \\
   & $\Lambda D^{-}$ & hadronic decay \\
$M_{\Theta_{cs}} < M_{\Lambda}+M_{D}$ & $\Lambda K^{+} \pi^{-}$,
$\Lambda K^{+} \pi^{+} \pi^{-} \pi^{-}$ & $ 0.41 \times 10^{-12}$ sec. \\
   & $\Lambda K^{+} \pi^{-} \pi^{-}$ & $ 1.0 \times 10^{-12}$ sec. \\
\hline
\end{tabular}
\end{center}
\label{table11}
\end{table}
\end{widetext}

\vspace{0.5cm}

Lastly, we comment on the possibility to measure doubly charmed
baryons in heavy ion collisions. The doubly charmed baryon
$\Xi_{cc}^{++}$ have been observed by the SELEX Collaboration in the
$\Lambda_{c}^{+}K^{-}\pi^{+}$ and in the $pD^{+}K^{-}$ decay modes
with a mass of ($3518.7 \pm 1.7$) MeV \cite{Mattson:2002vu,
Ocherashvili:2004hi}. The same collaboration has also
successfully measured $\Xi_{cc}^{+}$ in the
$\Lambda_{c}^{+}K^{-}\pi^{+}\pi^{+}$ decay mode with a mass of 3460
MeV \cite{Russ:2002bw}.  On the other hand, attempts by the FOCUS
Collaboration in the photoproduction experiment,
and by the BABAR in $e^{+}e^{-}$ annihilation experiments
\cite{Ratti:2003ez, Aubert:2006qw} have so far failed to
establish the existence of the doubly charmed baryons. Hence, it is
an interesting problem to search for the doubly charmed baryons in
heavy ion collisions. Using the coalescence model, we find that the
number of $\Xi_{cc}^{+}$ produced are $1.9 \times 10^{-5}$ at RHIC,
and $3.2 \times 10^{-4}$ at LHC. Therefore, we will be able to
realistically measure $\Xi_{cc}^{+}$ and $\Xi_{cc}^{++}$ through
their decay vertices to $\Lambda_{c}^{+}K^{-}\pi^{+}$ and
$pD^{+}K^{-}$, and to $\Lambda_{c}^{+}K^{-}\pi^{+}\pi^{+}$,
respectively.

\section{summary}

Based on the consideration of the color-spin interaction between
diquarks, which describes reasonably the mass splittings between
many hadrons and their spin flipped partners, we have shown that
tetraquark mesons and pentaquark baryons that consist of two charmed
quarks could be bound. Using the quark coalescence model, their
yields in heavy ion collisions at both RHIC and LHC are estimated.
Because of the expected large charm quark number in central Pb+Pb
collisions at LHC, the abundances of the tetraquark meson $T_{cc}$
and pentaquark baryon $\Theta_{cs}$ are about $10^{-4}$ and
$10^{-3}$, respectively. We have also discussed their decay modes to
illustrate how they can be identified in heavy ion collisions. In
our studies, we have not taken into account the hadronic effect on
the abundance of these charmed exotics, as hadronic reactions that
affect their annihilation and production are unknown. Since the
yields of $T_{cc}$ and $\Theta_{cs}$ from the coalescence model is
significantly smaller than those expected from the statistical
hadronization model, including the hadronic effect is expected to
increase their yields substantially and reduces the differences from
the predictions from the quark coalescence model and the statistical
hadronization model. Also, charmed hadrons would be more abundantly
produced, particularly the $T_{cc}$, if charm quarks are produced
from the QGP formed in these collisions. We also comment on
the possible measurement of doubly charmed
baryons in heavy ion collisions, 
and the estimated numbers 
are $1.9 \times 10^{-5}$ and $3.2 \times 10^{-4}$ at RHIC and
LHC, respectively. We thus expect that the open
and hidden charmed hadron physics will be an interesting subject in
the forthcoming heavy ion collision experiments.

\vspace{0,5cm}
\begin{acknowledgements} The work of SHL and SY was supported by
the Korea Research Foundation KRF-2006-C00011, while that of CMK and
WL was supported by the U.S. National Science Foundation under grant
No. PHY-0457265 and the Welch Foundation under grant No. A-1358.
\end{acknowledgements}

\end{document}